\documentclass[letter]{aa}  

\newcommand{\degg}{$^{\circ}$}
\newcommand{\tar}{VIK\,J2318$-$3113}

\usepackage{graphicx}
\usepackage{ulem}
\usepackage{txfonts}
\usepackage[]{hyperref}
\hypersetup{
    citecolor=cyan,
    colorlinks=true,
    linkcolor=red,
    filecolor=red,      
    urlcolor=magenta,
    pdftitle={J2318-3113},
    pdfpagemode=FullScreen,
    }
\begin{document}

   \title{VLBI observations of VIK J2318-3113, a quasar at z = 6.44}

   \subtitle{}

   \author{Y. Zhang\inst{1},
   T. An\inst{1,2,3},
   A. Wang\inst{1,3},
   S. Frey\inst{4,5},
   L. I. Gurvits\inst{6,7},
   K. \'E. Gab\'anyi\inst{8,9,4},
   K. Perger\inst{4}
    \and
   Z. Paragi\inst{6}
          }

   \institute{Shanghai Astronomical Observatory, Chinese Academy of Sciences, Nandan Road 80, Shanghai 200030, China\\
              \email{ykzhang@shao.ac.cn, antao@shao.ac.cn}
         \and
             Key Laboratory of Cognitive Radio and Information Processing, Guilin University of Electronic Technology, 541004 Guilin
         \and
             College of Astronomy and Space Sciences, University of Chinese Academy of Sciences, 19A Yuquanlu, Beijing 100049, China
        \and
             Konkoly Observatory, ELKH Research Centre for Astronomy and Earth Sciences, Konkoly Thege Mikl\'os \'ut 15-17, H-1121 Budapest, Hungary
        \and
            Institute of Physics, ELTE E\"otv\"os Lor\'and University, P\'azm\'any P\'eter s\'et\'any 1/A, H-1117 Budapest, Hungary
        \and
            Joint Institute for VLBI ERIC, Oude Hoogeveensedijk 4, 7991 PD Dwingeloo, The Netherlands
        \and
            Department of Astrodynamics and Space Missions, Delft University of Technology, Kluyverweg 1, 2629 HS Delft, The Netherlands
        \and
            Department of Astronomy, Institute of Geography and Earth Sciences, ELTE E\"otv\"os Lor\'and University, P\'azm\'any P\'eter s\'et\'any 1/A, H-1117 Budapest, Hungary
        \and
            ELKH-ELTE Extragalactic Astrophysics Research Group, E\"otv\"os Lor\'and University, P\'azm\'any P\'eter s\'et\'any 1/A, H-1117 Budapest, Hungary 
             }

   \date{Received Xxxx YY, 2022; accepted Zzzz WW, 2022}
    \authorrunning{Zhang et al.}
 
  \abstract
{The nature of jets in active galactic nuclei (AGNs) in the early Universe and their feedback to the host galaxy remains a highly topical question. Observations of the radio structure of high-redshift AGNs enabled by very long baseline interferometry (VLBI) provide indispensable input into studies of their properties and role in the galaxies' evolution. To date, only five AGNs at redshift $z>6$ have been studied with the VLBI technique.}
  { \tar\ is a recently discovered quasar at $z = 6.44$ that had not been imaged with VLBI before the current work. Here we present the first VLBI imaging results of this high-redshift quasar, with the aim of corroborating its high-resolution appearance with the physical model of the object.}
  {    We carried out VLBI phase-referencing observations of \tar\ using the Very Long Baseline Array at two frequencies, 1.6 and 4.7~GHz, and obtained the first view at the radio structure on the milliarcsecond scale. }
  {
  The source was clearly detected at 1.6~GHz. We found that almost all of its radio emission comes from the parsec-scale core region. Our dual-frequency observations constrain the spectral index and brightness temperature of the radio core. Its properties are similar to those of other known high-redshift radio-loud AGNs.}
  {}

   \maketitle
%

\section{Introduction}\label{sec:intro}
To date, about 200 quasars have been discovered at redshift $z>6$ \citep{2016ApJS..227...11B,2016ApJ...833..222J,2018PASJ...70S..35M}, of which the most distant  is located at $z= 7.64$ \citep{2021ApJ...907L...1W,2017FrASS...4....9P, 2020MNRAS.494..789R}. These luminous high-redshift quasars (HRQs) are found to host supermassive (up to $\ge$ 10$^{9}$ M$_{\odot}$) black holes (SMBHs) in their nuclei
\citep{2011Natur.474..616M,2015Natur.518..512W,2018Natur.553..473B}. However, the rapid growth of SMBHs during the first tenth of the cosmological history remains a great challenge for current theoretical models (see the review in \citealp{2020ARA&A..58...27I}).

The central SMBHs of these luminous HRQs are in a super-Eddington accretion state, and their high radiation pressure drives massive outflows that have a strong impact on the host galaxy environment \citep{2012ARA&A..50..455F}. On the other hand, radio jets can directly trace the activity and accretion state of an active galactic nucleus (AGN) \citep{2019ARA&A..57..467B}, 
which in turn can provide tight constraints on the fundamental radiative properties of the early SMBHs. However, jet-induced kinetic feedback has not received much attention in HRQs yet. This situation is partly due to the fact that jetted AGNs active in the radio domain account for only $\sim10\%$ of the total AGN population \citep{2016ApJ...831..168K}, and the fraction of known radio-emitting AGNs is  even smaller at high redshifts \citep{2019MNRAS.490.2542P}. The small high-redshift AGN samples are gradually increasing as more and more high-redshift quasars are discovered by large multi-band surveys, for example, in radio  \citep[e.g. with the Atacama Large Millimeter/sub-millimeter Array (ALMA),][]{2018ApJ...854...97D,2021ApJ...907L...1W};  optical and infrared (e.g. the Visible and Infrared Survey Telescope for Astronomy (VISTA) Kilo-degree Infrared Galaxy survey (VIKINGs), \citep{2013Msngr.154...32E}; the Panoramic Survey Telescope and Rapid Response System (Pan-STARRS) surveys, \citep{2016arXiv161205560C}); and X-ray bands \citep[e.g. the extended ROentgen Survey with an Imaging Telescope Array (eROSITA),][]{2020MNRAS.497.1842M,2021AstL...47..123K}. These observations are advancing the understanding of the co-evolution of the first-generation black holes in the Universe and their host galaxies.


Recently, \citet{2021A&A...647L..11I} reported a radio observational study of a quasar identified at $z = 6.44$, corresponding to an era close to the end of the re-ionisation. The source, named VIKING J231818.35$-$311346.3 (hereafter \tar), was discovered in the VIKING survey \citep{2013Msngr.154...32E} in the near-infrared band. Its redshift $z = 6.444 \pm 0.005$ was confirmed in the subsequent VLT/X-Shooter observation in the near-infrared band and the ALMA [C II] line survey in the sub-millimetre band \citep{2018ApJ...854...97D}. It has a bolometric luminosity of about $7.4 \times 10^{46}$ erg s$^{-1}$, suggesting the presence of a SMBH of over 
$6 \times 10^8 M_\odot$
\citep{2021A&A...647L..11I}. It has been detected by several radio telescopes, including the Australian Square Kilometre Array Pathfinder (ASKAP), the Australia Telescope Compact Array (ATCA), the upgraded Giant Metrewave Radio Telescope (uGMRT), and the Karl G. Jansky Very Large Array (JVLA) in the Very Large Array Sky Survey \citep[VLASS;][]{2020PASP..132c5001L}.~The source has a steep radio spectrum at frequencies from $\sim$ 1 GHz (flux density around 1 mJy) to 6 GHz (flux density around 0.1 mJy)  \citep[see Table~1 and Fig.~1 in][]{2022Ighina}.
Other shallower radio surveys, for instance the National Radio Astronomy Observatory (NRAO) Very Large Array (VLA) Sky Survey (NVSS), the Tata Institute of Fundamental Research Giant Metrewave Telescope
(GMRT) Sky Survey (TGSS), and the Sydney University Molonglo Sky Survey (SUMSS), did not detect it due to insufficient sensitivity. The source is one of the most distant $z>6$ radio-loud quasars discovered so far, with a radio loudness parameter $R$ (the ratio of the rest-frame 5~GHz radio luminosity to the  4400~\AA{} optical luminosity) between 31 and 70 \citep{2021A&A...647L..11I}. The lower limit of $R$ places it in the faintest regime of the high-redshift radio-loud quasars \citep{2021ApJ...909...80B}. Comparison of ASKAP observations at multiple epochs reveals possible variability, which could be  intrinsic or caused by refractive interstellar scintillation \citep{2022Ighina}.
It has a steep spectrum with a power-law slope of $\alpha = -1.24$ between 888~MHz and 5~GHz, corresponding to the rest-frame range $7-40$~GHz (here the spectral index $\alpha$ is defined under the convention $S \propto \nu^{\alpha}$, where $S$ is the flux density and $\nu$ the observing frequency). The measurement with uGMRT at 400~MHz shows a tendency for the spectrum to flatten below 800~MHz \citep[see Fig.~1 in][]{2022Ighina}. The radio properties are consistent with those of a young radio jet \citep[][]{2022Ighina}.

To confirm the nature of the radio emission from this HRQ, and in particular to check whether relativistic beaming has a significant effect on the radio emission, we have carried out high-resolution very long baseline interferometry (VLBI) observations of \tar. The VLBI images allow the direct study of radio emission from the area surrounding galactic nuclei, excluding contamination from  star-forming activity on galactic scales \citep{2020ApJ...905L..32F}. Moreover, the Doppler-boosting factor can be estimated directly from VLBI observations.
The paper is organised as follows. Section~\ref{sec:obs} gives the VLBI observation and data reduction details of \tar. Section~\ref{sec:res} presents the high-resolution images and observational results of \tar\ and discusses the radio properties of this source. Section~\ref{sec:sum} summarises the paper and gives some perspectives on further radio studies of such high-$z$ sources. Throughout this paper the Lambda cold dark matter ($\Lambda$CDM) cosmological model parameters of $H_{0}=70$~km~\,s$^{-1}$\, Mpc$^{-1}$, $\Omega_\mathrm{m} = 0.3$, and $\Omega_{\Lambda} = 0.7$ are applied. At $z = 6.44$, 1 milliarcsecond (mas) angular size corresponds to a projected linear scale of $5.49$~pc \citep{2006PASP..118.1711W}.

\section{Observation and data reduction} \label{sec:obs}
We used the Very Long Baseline Array (VLBA) of NRAO to observe \tar\ at 4.7 GHz (6 cm) and 1.6 GHz (20 cm) (project code: BZ083). The observations were carried out on 2021 August 2 at 4.7 GHz and 2021 August 16 at 1.6 GHz. As the target is located in the southern hemisphere at fairly low declination (Dec $<-30$\degg), the participating antennas and the observation time range were carefully selected to ensure that as many antennas as possible could be used (i.e.  source elevation angles  higher than 15\degg) during the observation period. In the end, seven of the ten VLBA antennas were selected (excluding the northernmost Brewster and the Hancock and North Liberty telescopes in the east of the US). The total flux density of this source is only about 1~mJy at $1-5$~GHz \citep{2022Ighina}, so phase-referenced observations \citep{1995ASPC...82..327B} were performed, lasting for 5~h in each frequency band. The a priori coordinates of \tar\ were taken from the ALMA detection at 247~GHz:  right ascension RA = $23^\mathrm{h} 18^\mathrm{m} 18.351^\mathrm{s}$, declination Dec = $-31^{\circ} 13^{\prime} 46.350^{\prime\prime}$ (J2000). The bright source J2314$-$3138 ($S \approx 500$~mJy) was used as the phase-reference calibrator, which is located at RA = $23^\mathrm{h} 14^\mathrm{h} 48.5006^\mathrm{s}$, Dec = $-31^{\circ} 38^{\prime} 39.5264^{\prime\prime}$ (J2000) (from the Astrogeo astrometric database\footnote{\url{http://astrogeo.org}}) and only 0.85\degg\ away from the target source. The positional accuracy of J2314$-$3138 is 0.13~mas. The duty-cycle for the phase-referencing observations was 5~min, including about 4~min on the target source and 1~min on the 
calibrator. The recording data rate was 2~Gbps. Other observational information is listed in Table~\ref{tab:obs}. 

The raw data recorded at each telescope were correlated at the Distributed FX (DiFX) correlator \citep{2011PASP..123..275D} in Socorro (New Mexico, USA). The output data integration time was set for 2~s.
The correlated data were downloaded to the China Square Kilometre Array (SKA) Regional Centre \citep{2019NatAs...3.1030A} for further calibration and imaging using the NRAO Astronomical Image Processing System (AIPS) software package \citep{2003ASSL..285..109G}.
We developed a pipeline (Lao et al. in prep.) based on the standard calibration procedure of the AIPS Cookbook\footnote{\url{http://www.aips.nrao.edu/cook.html}}, with the following main processing steps. 
We first inspected the data quality using the procedure \textsc{vlbasumm} and the task \textsc{possm}. Based on the results from data quality check, we used Kitt Peak (KP) as the reference antenna. After the correction for Earth orientation and atmosphere parameters using the task \textsc{clcor}, we used \textsc{apcal} to calibrate the visibility amplitudes. During the amplitude calibration, the weather information and antenna parameters were considered to better calibrate the gain factors and opacity effects for each antenna.
Next, we chose an appropriate scan on the bright fringe-finding calibrator 3C\,454.3 to correct the instrumental delay and phase errors, and applied the solutions to all data.
Then we ran fringe fitting on the phase-referencing calibrator J2314$-$3118 using the task \textsc{fring} to solve for the global phase errors. The derived gain solutions were applied to all visibilities. 
Finally, the antenna-based bandpass functions were determined based on 3C~454.3 by using the task \textsc{bpass} and applied to all data. 

\begin{figure*}[ht]
\centering
\includegraphics[width=0.3\textwidth]{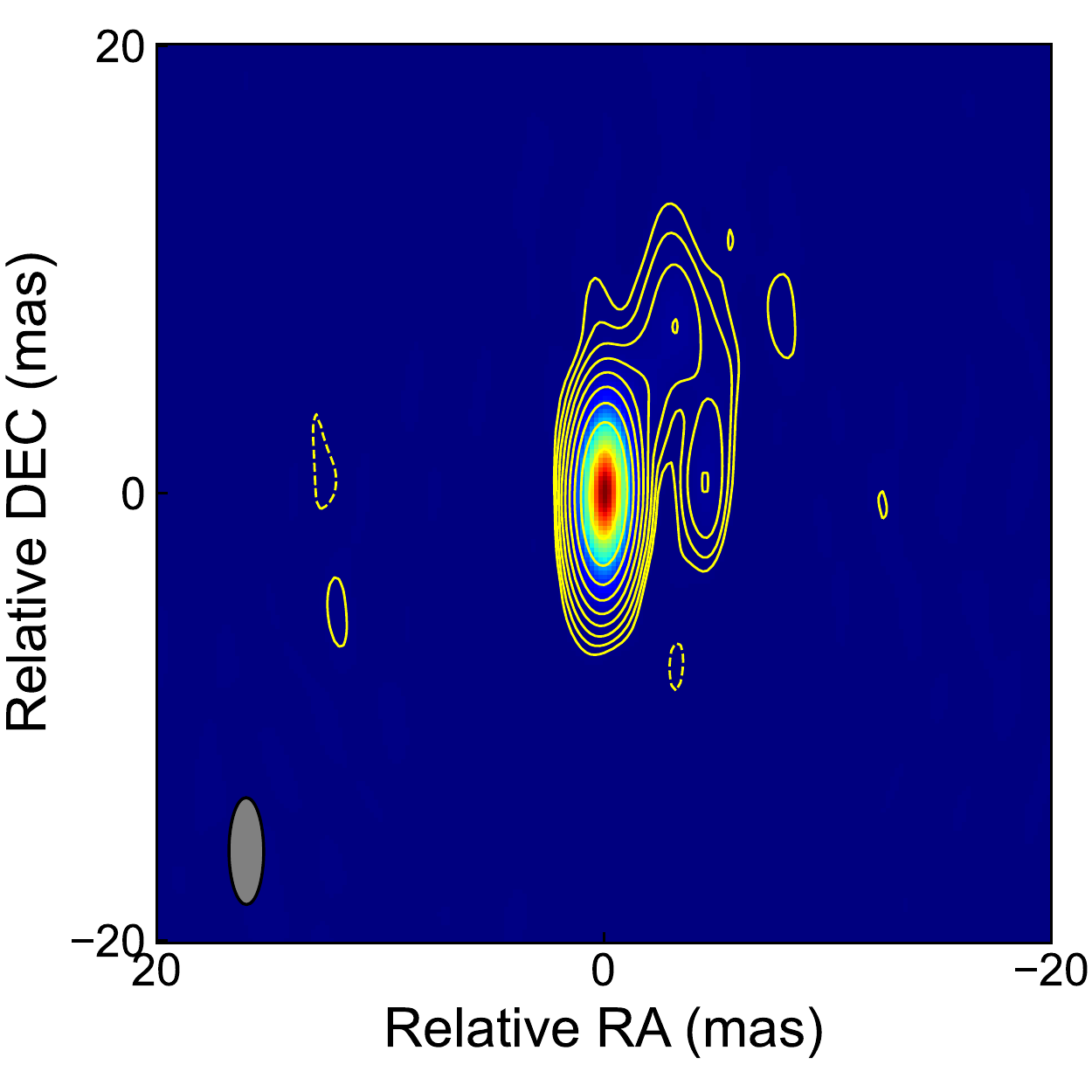}
\includegraphics[width=0.3\textwidth]{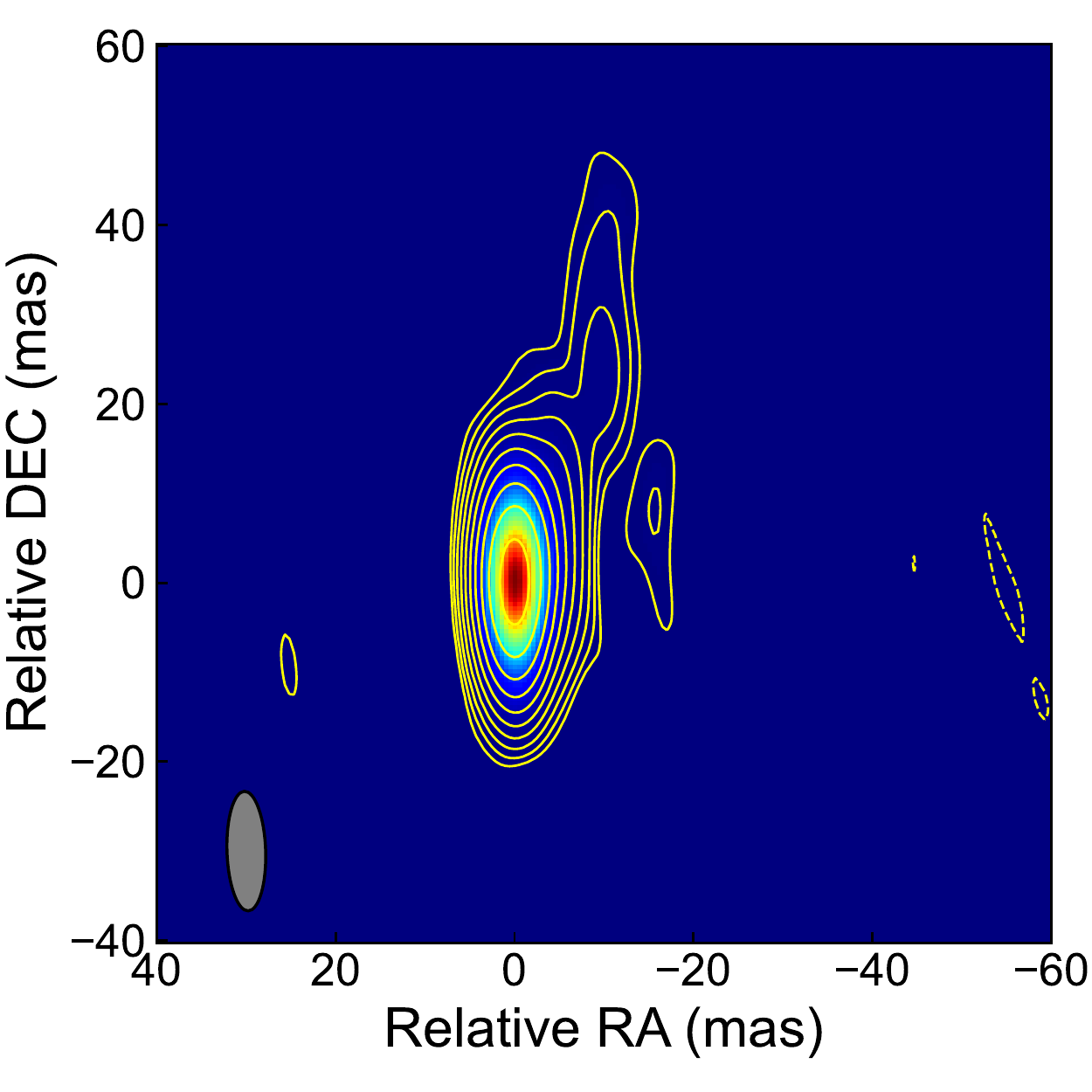}
\caption{Naturally weighted images of the phase-reference calibrator J2314$-$3138 after self-calibration. In each image the lowest contours represent $\pm 5$ times the root mean square (rms) noise and the positive contours increment by a factor of 2. Left:  4.7-GHz VLBA image of J2314$-$3138 from session BZ083A observed on 2021 August 2. The peak intensity is 282~mJy\,beam$^{-1}$, the rms noise is $\sigma=0.14$~mJy\,beam$^{-1}$. The half-power width of the elliptical Gaussian restoring beam is 4.8~mas $\times$ 1.6~mas with a major axis position angle $\textrm{PA}=0.2$\degg, as presented in the bottom left corner. Right:  1.6-GHz VLBA image of J2314$-$3138 from session BZ083B observed on 2021 August 16. The peak intensity is 533~mJy\,beam$^{-1}$, $\sigma=0.16$~mJy\,beam$^{-1}$. The restoring beam is 13.4~mas $\times$ 4.3~mas with $\textrm{PA}=1.9$\degg. \label{fig:pref}}
\end{figure*}

The calibrated visibility data of the phase calibrator, J2314$-$3138, were first exported as an external single-source FITS file, averaged over each 128 MHz  sub-band and 2 s sampling time. Then these data were loaded into the {\sc Difmap} software \citep{1994BAAS...26..987S} for the final self-calibration and imaging. 
This process consists of two steps on the imaging. The first step conducts \textsc{clean} and phase-only self-calibration algorithms iteratively to subtract bright data points until no robust peaks (e.g. brighter than five times   the background noise) can be found in the residual map. In the second step, the phase-and-amplitude self-calibration is applied to the previous iterations, where the solution intervals start with a few hours and end with 4~min, reducing by a factor of two in each iteration \citep[see][]{1994BAAS...26..987S,1984ARA&A..22...97P}.
This  corrects the complex antenna gains against time and makes the deconvolved image more robust.
The final \textsc{clean} images of J2314$-$3138 are presented in the Fig.~\ref{fig:pref}.
Then the images of the calibrator were imported back to AIPS, to run fringe fitting again by using the \textsc{clean} components as the input model. This can correct the phase offsets caused by the intrinsic structure of the phase-reference calibrator itself.
Finally, after applying all the calibration solutions to the target source, the visibility data of \tar\ were exported and imaged in {\sc Difmap}. Due to the weakness of \tar, no self-calibration was performed.

\begin{table*} 
\caption{Observational set-up of the VLBA sessions on \tar. \label{tab:obs}}
\centering
\begin{tabular}{cccccc}
\hline\hline     
Session & Date  & $\nu_\mathrm{obs}$ & Time & Bandwidth & Antennas \\
        &       & (GHz)& (h) & (MHz) & \\
(1) & (2) & (3) & (4) & (5) & (6) \\    
\hline
BZ083A          & 2021 Aug 02                   &  4.74   &  5  &  512 MHz  & SC, FD, LA,  PT, KP, OV, MK \\
BZ083B          & 2021 Aug 16                   &  1.57   &  5  &  512 MHz  & SC, FD, LA, PT, KP, OV, MK \\
\hline
\end{tabular}\\

Col.~1: Observation ID; Col.~2: Observation date; Col.~3: Central frequency; Col.~4: Total observation time; Col.~5: Data recording bandwidth; Col.~6:  VLBA telescopes that participated in these observations:  FD (Fort Davis), KP (Kitt Peak), LA (Los Alamos), MK (Mauna Kea), OV (Owens Valley), PT (Pie Town), and SC (Saint Croix).
\end{table*}

\section{Results and discussion} \label{sec:res}
Our VLBA observations allowed us to detect the milliarcsecond-scale radio morphology as well as the precise location of the radio-emitting region of this high-redshift quasar. Combined with the historical radio data from ASKAP and VLASS, we were able to constrain the compactness and the radio classification of \tar.

\begin{figure*}
\includegraphics[width=0.45\textwidth]{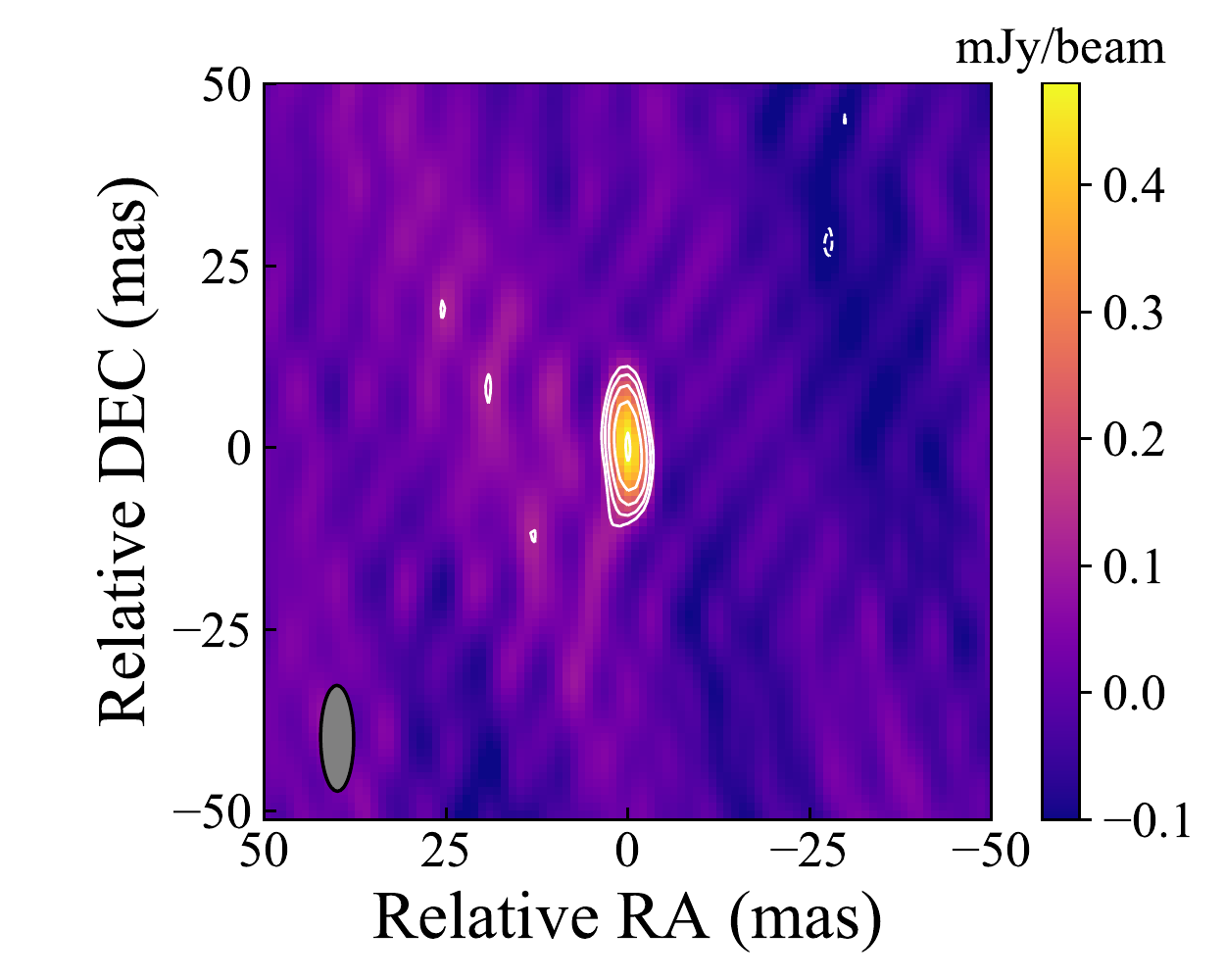}
\includegraphics[width=0.45\textwidth]{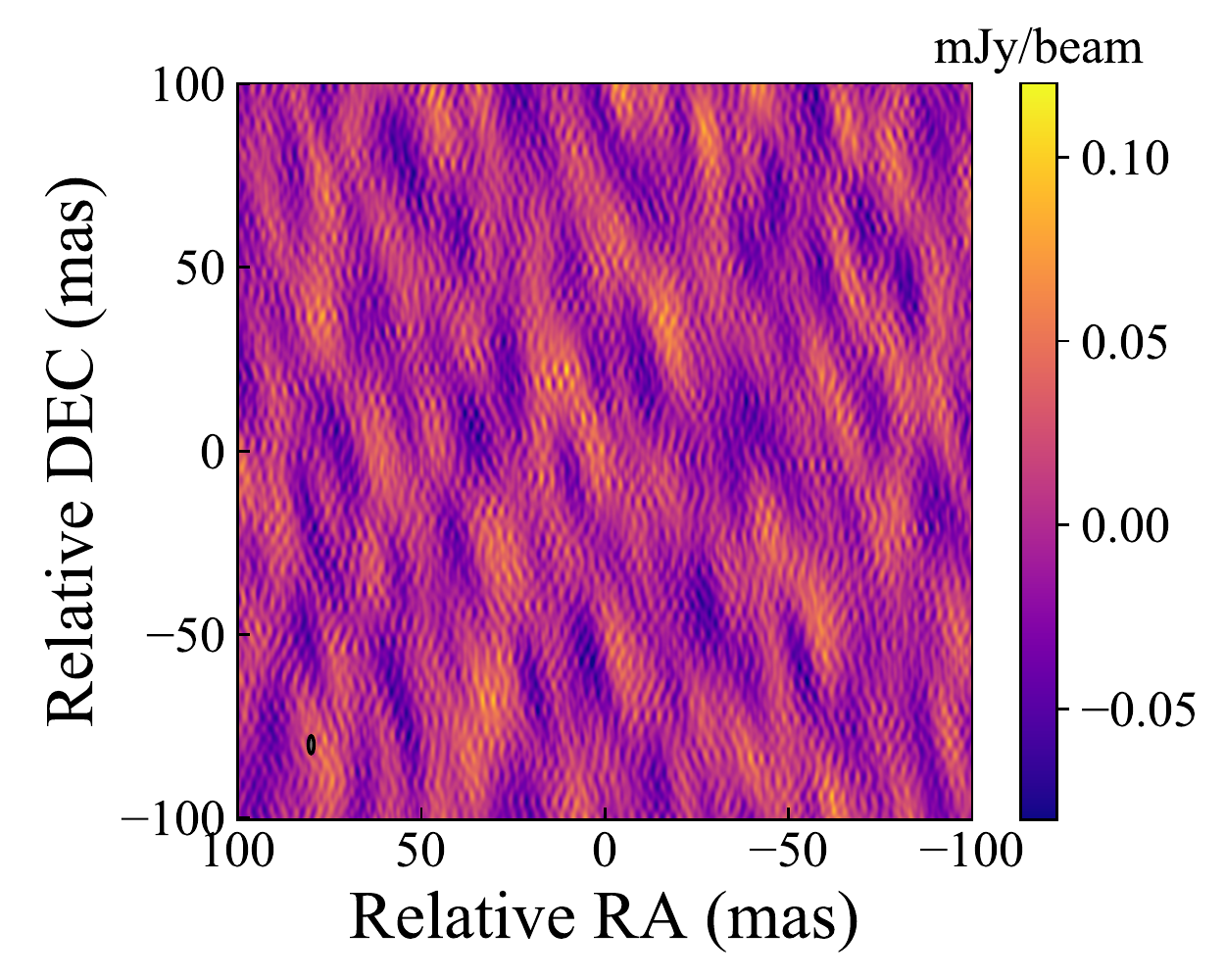}
\caption{Naturally weighted \textsc{clean} images of \tar\ made from the VLBA observations. Detailed parameters are listed in Table~\ref{tab:img}. Left: Target source detected as a compact component at 1.57~GHz, with $\mathrm{S/N} \sim 12$. 
The contour levels are at ($-1$,1,$\sqrt{2}$,2,2$\sqrt{2}$,4,4$\sqrt{2}$) $\times$ the 3$\sigma$ image noise.
Right: Non-detection dirty map of the target source at 4.74~GHz, where no prominent peaks above $0.13$~mJy\,beam$^{-1}$ ($5\sigma$) could be found. The image centre is the same as in the 1.6 GHz image.
\label{fig:tar}}
\end{figure*}

The VLBA images of \tar~are presented in Fig.~\ref{fig:tar}, where we can see that the source is detected at 1.6~GHz, but not at 4.7~GHz.
The detection at 1.6~GHz shows a compact structure with a signal-to-noise ratio (S/N) of $\sim$ 12. We fitted the target visibility data using a circular Gaussian brightness distribution model in \textsc{Difmap}. The corresponding flux density is 0.55~mJy, and the fitted component size is 2.3~mas (full width at half maximum, FWHM). The fitted component is close to the minimum resolvable size with the interferometer array (2.2~mas) \citep{2005AJ....130.2473K} and the \textsc{clean} map exhibits a compact morphology reflecting the shape of the restoring beam.

The source peak position is at RA = $23^\mathrm{h}  18^\mathrm{m} 18.3684^\mathrm{s}$ (with an estimated uncertainty of $0.0004^{\prime\prime}$), Dec=$-31^{\circ} 13^{\prime} 46.4030^{\prime\prime}$ (uncertainty $0.0008^{\prime\prime}$). 
The positional errors are a combination of the statistical error in the image domain ($\frac{\phi_{b}}{2(\mathrm{S/N})}$, where $\phi_{b}$ is the beam size projected along the  RA or Dec directions), the position error of the phase-reference calibrator J2314$-$3138 (from the latest measurements in Astrogeo), and the astrometric error related to the offset between the target source and the phase calibrator ($\sim 0.7$\degg in RA and $\sim 0.4$\degg in Dec). From the VLBI astrometric accuracy simulation for extremely low declinations \citep{2006A&A...452.1099P}, we used the extreme-case value of the simulated errors as our astrometric errors, with 0.3~mas along the RA and 0.5~mas along the  Dec directions. Our VLBI measurement led to the first precise position determination of the radio core of the high-redshift source \tar.

At 4.7 GHz, no detectable signal peak  (i.e. $\ge 5 \sigma$) was found at the same coordinates as the 1.6 GHz detection. Recent ATCA measurements at 5~GHz indicated a flux density of 0.1~mJy \citep{2022Ighina}, corresponding to approximately $3.8\sigma$ in our observation, if we assume that all the ATCA flux density can be detected by VLBA. Therefore, a higher-sensitivity VLBI observation is necessary in the future to explore the milliarcsecond-scale compact source structure at 5~GHz. Using the 5$\sigma$ value as the 5 GHz flux density upper limit, and applying the minimum resolvable size based on the VLBI observing parameters \citep{2005AJ....130.2473K}, an upper limit of the radio spectral index $\alpha$ of the compact structure, and a lower limit of the brightness temperature can be obtained.
Table~\ref{tab:img} lists the above observing and model-fitting parameters of \tar.
The errors of the VLBI parameters are estimated through applying the formulae from \citet{1999ASPC..180..301F}.
Taking the 5$\sigma$ value (i.e. $S_\mathrm{4.7} \le 0.13$~mJy) as the upper limit of the source flux density at 4.7~GHz, the corresponding spectral index is $\alpha_{1.6}^{4.7} \le -1.2$.
Compared with the multi-frequency result obtained from low-resolution observations \citep[e.g. ATCA, VLA, ASKAP; see][]{2022Ighina}, the steep spectrum we found with $\alpha_{1.6}^{4.7} \le -1.2$ is in good agreement. The spectral index derived by \citet{2022Ighina}, fitted value $\alpha_\mathrm{fit} \sim -1.5$ and measured value $\alpha_\mathrm{meas} \sim -1.2$, is characteristic for the whole source. Based on the flux densities at much smaller angular scales, our result shows similar spectral properties for the highly compact radio-emitting region confined to $\le 2.3$~mas. At 1.6~GHz, our radio brightness detected with the VLBA is almost equal to that obtained from the ASKAP mid-band survey, $0.5 \pm 0.2$~mJy\,beam$^{-1}$ at 1.4~GHz. It suggests that almost all radio emission comes from the compact core, leaving little room for galactic-scale star formation in the quasar host galaxy to contribute to the gigahertz frequency radio emission.  

The brightness temperature ($T_\mathrm{B}$) of the detected compact component is estimated as follows \citep[e.g. ][]{1982ApJ...252..102C}:

\begin{equation} \label{equ:tb}
T_\mathrm{B}=1.22 \times 10^{12} (1+z) \frac{S_{\nu}}{\theta_\mathrm{comp}^{2}\nu^{2}} \, \mathrm {K}.
\end{equation}
Here S$_{\nu}$ is the flux density of the VLBI component (measured in Jy), $\nu$ the observing frequency in GHz, and  $\theta_\mathrm{comp}$  the diameter (FWHM) of the circular Gaussian model component (in mas). If $\theta_\mathrm{comp}$ is an upper limit, then the resulting $T_\mathrm{B}$ is a lower limit. The values derived from our VLBA observations are in Table~\ref{tab:img}. Based on the high brightness temperature ($\sim 10^{8}$~K) and compact emission zone, we confirm that the radio emission of the source is dominated by AGN activity. This could also be proved by using the wide-band spectral energy distribution templates from quasars and starburst galaxies, where the radio emission is $1-2$ orders of magnitude higher than the predicted star formation-dominated radio emission \citep{2022Ighina}.
Taking advantage of the spectral index constraints, the VLBI observation yields a lower limit of the monochromatic radio luminosity at 1.6~GHz as $P \geq 6.3 \pm 0.8 \times 10^{25}$~W\,Hz$^{-1}$.

\begin{table*}
\caption{Image parameters and the estimated brightness temperatures of \tar. \label{tab:img}}
                \centering
        \begin{tabular}{ccccccccc}
            \hline\hline
            $\nu$       & $S_\mathrm{peak}$     & $S_\mathrm{model}$ & $\sigma$  & $\theta_\mathrm{comp}$ & $B_\mathrm{maj}$ & $B_\mathrm{min}$ & $B_\mathrm{PA}$  & $T_\mathrm{B}$ \\
            (GHz)& (mJy\,bm$^{-1}$)     & (mJy)& (mJy\,bm$^{-1}$) & (mas) &  (mas)& (mas)& (\degg) & ($\times$10$^{8}$K)\\
            \hline
            1.57        & 0.48$\pm$0.05 & 0.55$\pm$0.07  & 0.039 & 2.3$\pm$0.7 & 14.6 & 4.6  & 0.3     & 4.0$\pm$1.7\\
            4.74        & $<0.13$&  ...   & 0.026 &$\le$1.2 &4.8 & 1.6 & $-$0.3 & $\ge$0.4   \\
            \hline
        \end{tabular}\\
    Notes: Col.~1 -- Observing frequency; Col.~2 -- Peak intensity of the image; Col.~3 -- Total flux density of the fitted model component of the emitting region; Col.~4 -- R.m.s noise of the image background; Col.~5 -- FWHM size of the fitted component; Cols.~6--7 -- FWHM size along the major and minor axis of the restoring beam; Col.~8 -- Position angle of the beam major axis, measured from north to east; Col.~9 -- Estimated brightness temperature of the fitted component based on Eq.~\ref{equ:tb}. When the fitted size is an upper limit, then the brightness temperature could be a lower limit (if the flux density at 4.7 GHz equals its upper limit).
\end{table*}

Advances in the study of HRQs have attracted attention from both observational and theoretical cosmology standpoints, helping to unravel the mysteries of early evolution of the Universe. In popular cosmological models, cosmic reionisation ends around the epoch $z = 7$ \citep{2001PhR...349..125B}. Current instrumental capabilities enable observations reaching the threshold of the hydrogen reionisation epoch, and the James Webb Space Telescope will penetrate deeper into the more distant Universe \citep{2006SSRv..123..485G}.

The recent discovery of billion solar mass SMBHs in luminous $z > 7$ quasars at the epoch of reionisation poses the most stringent constraints on the masses of the seed black holes and the early accreting mode \citep{2020ApJ...897L..14Y,2021ApJ...908...53W}. Jets are closely related to accretion and AGN feedback of SMBHs, but no radio jets have been found so far in $z>7$ quasars, and continued exploration is still needed. 
High-redshift jets are difficult to grow to large sizes \citep{2014MNRAS.438.2694G}, so observing the jets of HRQs requires very high resolution, and VLBI is typically used to probe such radio jet structures.
Among the more than 200 quasars at $z>6$ that have been discovered so far, only five others have been observed by VLBI:  
J0309$+$2717 ($z=6.10$, \citealp{2020A&A...643L..12S}),
J1427$+$3312 ($z=6.12$, \citealp{2008A&A...484L..39F,2008AJ....136..344M}),
J1129$+$1846 ($z=6.82$, \citealp{2021AJ....161..207M}),
J1429$+$5447 ($z=6.21$, \citealp{2011A&A...531L...5F}),
and J0100$+$2802 ($z=6.33$, \citealp{2017ApJ...835L..20W}).
Among the VLBI detected HRQs, J0100$+$2802 is the only radio-quiet source with $R < 10$. J0309$+$2717 (PSO J030947.49+271757.31) is the only blazar with $\sim 20$~mJy flux density at 1.5~GHz \citep{2020A&A...643L..12S}. The flux density of the others is at mJy and sub-mJy levels.  \tar\ presented in this paper and J2331+1129 (S. Frey et al., in prep.) add two $z>6$ jetted quasars which also have mJy flux densities in VLBI images. The general picture emerging from their VLBI imaging observations is that the radio-loud HRQs usually have compact but somewhat resolved structures, including one (J1427$+$3312) which is a double \citep{2008A&A...484L..39F,2008AJ....136..344M}. Except J0309$+$2717, these $z>6$ quasars showed steep spectra in the (observed) gigahertz frequency range, and are identified as nascent radio quasars or a compact symmetric object, which could results from the GPS (Gigahertz-Peaked Spectrum) or MPS (Megahertz-Peaked Spectrum) source nature \citep[e.g.][]{2021A&ARv..29....3O}. 

The rate of detection of jets in HRQs should be treated with caution due to a combined effect of instrumental and astrophysical selection. On the one hand, most VLBI studies of HRQ objects are conducted at centimetre wavelengths, which correspond to $(1+z)$ times shorter wavelengths in the source rest frame. However, jets have steep spectra, thus their detection at short wavelengths is becoming more difficult with growing redshift. This is an astrophysical factor of selection that works toward the appearance of barely resolved or unresolved structures in HRQs \citep{Gurvits+2000,2015IAUS..313..327G}. On the other hand, detecting jets at their `usual' centimetre wavelengths (gigahertz frequencies) in the source rest frames requires  milliarcsecond resolution at observing wavelengths $(1+z)$ times longer, that is, many decimetres or even metres. However, the limiting factor in this case is the size of Earth: even global VLBI systems cannot reach the required angular resolution at these wavelengths. Baselines larger than the Earth's diameter   need to be exploited, that is, observing with Space VLBI \citep{2020AdSpR..65..850A,2020AdSpR..65..868G}.

Assuming that the observed viewing angles of high-redshift radio-loud quasars are not essentially different from those of low-redshift quasars, the reason why most high-$z$ quasars have systematically lower radio core bright temperatures and radio loudness than low-redshift quasars should be intrinsic. High-redshift jetted AGNs are observed to host massive black holes;  major mergers in the early Universe would trigger fast spinning of the black holes, but these black holes do not seem to produce highly relativistic jets. One possibility is that the central engines in high-redshift AGNs have not yet reached the magnetically arrested disk (MAD) state \citep{2003PASJ...55L..69N}. A similar case can be seen in the nearby radio-intermediate quasar III~Zw~2 \citep{2021A&A...652A..14C}. Measurements of the magnetic field strength in high-$z$ jets and high-resolution numerical simulations would help to solve this puzzle. 

\section{Conclusions}\label{sec:sum}

We present the first VLBI imaging results of the high-redshift radio-loud quasar \tar~using 1.6~GHz (5 cm) and 4.7~GHz (20 cm) VLBA observations. Precise coordinates and VLBI parameters were obtained for this source.
The 1.6 GHz image shows a compact core with a modelled size of $2.3$~mas (FWHM) and flux density of $5.5$~mJy. At 4.7 GHz the source remains undetected due to its steep spectrum and the insufficient sensitivity.
Our VLBI data alone constrain the spectral index as $\alpha_{1.6}^{4.7} \le -1.2$, consistently with the spectral index obtained for the whole source \citep{2022Ighina}. This is not surprising because the radio emission is dominated by the parsec-scale VLBI component.
The brightness temperature $T_\mathrm{B}$ from the detection at 1.6 GHz is $\sim 4 \times 10^8$~K, confirming the AGN origin of the radio emission.

Our results show good consistency with the multi-band radio study from low-resolution observations \citep{2022Ighina}. A very high VLBI dominance (i.e. $S_{\rm VLBI}/S_{\rm ATCA}\ge 70\%$) was also obtained by considering the external variability introduced by  refractive interstellar scintillation \citep[see][]{2022Ighina}. Combining all the factors above, our work established that the radio emission of \tar\ mainly comes from the innermost region ($\le 2.2$~mas, corresponding to $12$~pc linear size), originating from the young radio jets formed at around the end of the reionisation epoch in the history of the Universe.

Notably, this is the first VLBI detection of a $z>6$ radio-loud AGN having a fairly low declination in the southern hemisphere. With the development of SKA precursors (e.g. the Murchison Widefield Array (MWA) \citep{2013PASA...30....7T}, ASKAP, and the MeerKAT radio telescope \citep{2016mks..confE...1J}), these results would provide important material for the study of the AGN evolution during the cosmic dawn in the SKA era.

\begin{acknowledgements}
This research was funded by National Key R\&D Programme of China under grant number 2018YFA0404603 and the China--Hungary project funded by the the Ministry of Science and Technology of China and Chinese Academy of Sciences. YKZ was sponsored by Shanghai Sailing Program under grant number 22YF1456100. This research was supported by the Hungarian National Research, Development and Innovation Office (NKFIH), grant number OTKA K134213.
The National Radio Astronomy Observatory is a facility of the National Science Foundation operated under cooperative agreement by Associated Universities, Inc. We acknowledge the use of data from the Astrogeo Center database maintained by Leonid Petrov.
\end{acknowledgements}


\begin{thebibliography}{999}
        \bibitem[An et al.(2019)]{2019NatAs...3.1030A} An, T., Wu, X.-P., \& Hong, X.\ 2019, Nature Astronomy, 3, 1030.
        
        \bibitem[An et al.(2020)]{2020AdSpR..65..850A} An, T., Hong, X., Zheng, W., et al.\ 2020, Advances in Space Research, 65, 850.
        
        \bibitem[Barkana \& Loeb(2001)]{2001PhR...349..125B} Barkana, R., \& Loeb, A.\ 2001, \physrep, 349, 125.
        
        \bibitem[Ba{\~n}ados et al.(2016)]{2016ApJS..227...11B} Ba{\~n}ados, E., Venemans, B.~P., Decarli, R., et al.\ 2016, \apjs, 227, 11.
        
        \bibitem[Ba{\~n}ados et al.(2018)]{2018Natur.553..473B} Ba{\~n}ados, E., Venemans, B.~P., Mazzucchelli, C., et al.\ 2018, \nat, 553, 473.
        
        \bibitem[Ba{\~n}ados et al.(2021)]{2021ApJ...909...80B} Ba{\~n}ados, E., Mazzucchelli, C., Momjian, E., et al.\ 2021, \apj, 909, 80.
        
        \bibitem[Beasley \& Conway(1995)]{1995ASPC...82..327B} Beasley, A.~J., \& Conway, J.~E.\ 1995, Very Long Baseline Interferometry and the VLBA, 82, 327.
        
        \bibitem[Blandford et al.(2019)]{2019ARA&A..57..467B} Blandford, R., Meier, D., \& Readhead, A.\ 2019, \araa, 57, 467.
        
        \bibitem[Chamani et al.(2021)]{2021A&A...652A..14C} Chamani, W., Savolainen, T., Hada, K., \& Xu, M.~H.\ 2021, \aap, 652, A14.
        
        \bibitem[Chambers et al.(2016)]{2016arXiv161205560C} Chambers, K.~C., Magnier, E.~A., Metcalfe, N., et al.\ 2016, arXiv e-prints, arXiv:1612.05560.
        
        \bibitem[Condon et al.(1982)]{1982ApJ...252..102C} Condon, J.~J., Condon, M.~A., Gisler, G., \& Puschell, J.~J.\ 1982, \apj, 252, 102.
        
        \bibitem[Decarli et al.(2018)]{2018ApJ...854...97D} Decarli, R., Walter, F., Venemans, B.~P., et al.\ 2018, \apj, 854, 97.
        
        \bibitem[Deller et al.(2011)]{2011PASP..123..275D} Deller, A.~T., Brisken, W.~F., Phillips, C.~J., et al.\ 2011, \pasp, 123, 275.
        
        \bibitem[Edge et al.(2013)]{2013Msngr.154...32E} Edge, A., Sutherland, W., Kuijken, K., et al.\ 2013, The Messenger, 154, 32.
        
        \bibitem[Fabian(2012)]{2012ARA&A..50..455F} Fabian, A.~C.\ 2012, \araa, 50, 455.
        
        \bibitem[Fan et al.(2020)]{2020ApJ...905L..32F} Fan, L., Chen, W., An, T., et al.\ 2020, \apjl, 905, L32.
        
        \bibitem[Fomalont(1999)]{1999ASPC..180..301F} Fomalont, E.~B.\ 1999, Synthesis Imaging in Radio Astronomy II, 180, 301.
        
        \bibitem[Frey et al.(2008)]{2008A&A...484L..39F} Frey, S., Gurvits, L.~I., Paragi, Z., \& {\'E}. Gab{\'a}nyi, K.\ 2008, \aap, 484, L39.
        
        \bibitem[Frey et al.(2011)]{2011A&A...531L...5F} Frey, S., Paragi, Z., Gurvits, L.~I., Gab{\'a}nyi, K. {\'E}., \& Cseh, D.\ 2011, \aap, 531, L5.
        
        \bibitem[Gardner et al.(2006)]{2006SSRv..123..485G} Gardner, J.~P., Mather, J.~C., Clampin, M., et al.\ 2006, \ssr, 123, 485.
        
        \bibitem[Ghisellini et al.(2014)]{2014MNRAS.438.2694G} Ghisellini, G., Celotti, A., Tavecchio, F., Haardt, F., \& Sbarrato, T.\ 2014, \mnras, 438, 2694.
        
        \bibitem[Greisen(2003)]{2003ASSL..285..109G} Greisen, E.~W.\ 2003, Information Handling in Astronomy - Historical Vistas, 285, 109.
        
        \bibitem[Gurvits(2000)]{Gurvits+2000} Gurvits, L.I.\ 2000, in Perspectives on Radio Astronomy: Science with Large Antenna Arrays, ed. van Haarlem, M.~P., Dwingeloo:ASTRON, 183.
        
        \bibitem[Gurvits(2020)]{2020AdSpR..65..868G}Gurvits, L.I.: 2020, {\it Advances in Space Research} {\bf 65}, 868. 
        
        \bibitem[Gurvits, Frey, and Paragi(2015)]{2015IAUS..313..327G} Gurvits, L.I., Frey, S., and Paragi, Z.: 2015, {\it Extragalactic Jets from Every Angle} {\bf 313}, 327. 
        
        \bibitem[Ighina et al.(2021)]{2021A&A...647L..11I} Ighina, L., Belladitta, S., Caccianiga, A., et al.\ 2021, \aap, 647, L11.
        
        \bibitem[Ighina et al.(2022)]{2022Ighina} Ighina, L., Leung, J.~K., Broderick, J.~W., et al.\ 2022, arXiv e-prints, arXiv:2203.08142.
        
        \bibitem[Inayoshi et al.(2020)]{2020ARA&A..58...27I} Inayoshi, K., Visbal, E., \& Haiman, Z.\ 2020, \araa, 58, 27.
        
        \bibitem[Jiang et al.(2016)]{2016ApJ...833..222J} Jiang, L., McGreer, I.~D., Fan, X., et al.\ 2016, \apj, 833, 222.
        
        \bibitem[Jonas \& MeerKAT Team(2016)]{2016mks..confE...1J} Jonas, J. \& MeerKAT Team\ 2016, MeerKAT Science: On the Pathway to the SKA, 1
        
        \bibitem[Kellermann et al.(2016)]{2016ApJ...831..168K} Kellermann, K.~I., Condon, J.~J., Kimball, A.~E., Perley, R.~A., \& Ivezi{\'c}, {\v{Z}}.\ 2016, \apj, 831, 168.
        
        \bibitem[Khorunzhev et al.(2021)]{2021AstL...47..123K} Khorunzhev, G.~A., Meshcheryakov, A.~V., Medvedev, P.~S., et al.\ 2021, Astronomy Letters, 47, 123.
        
        \bibitem[Kovalev et al.(2005)]{2005AJ....130.2473K} Kovalev, Y.~Y., Kellermann, K.~I., Lister, M.~L., et al.\ 2005, \aj, 130, 2473.
        
        \bibitem[Lacy et al.(2020)]{2020PASP..132c5001L} Lacy, M., Baum, S.~A., Chandler, C.~J., et al.\ 2020, \pasp, 132, 035001.
        
        
        \bibitem[Maini et al.(2016)]{2016A&A...589L...3M} Maini, A., Prandoni, I., Norris, R.~P., Giovannini, G., \& Spitler, L.~R.\ 2016, \aap, 589, L3.
        
        \bibitem[Matsuoka et al.(2018)]{2018PASJ...70S..35M} Matsuoka, Y., Onoue, M., Kashikawa, N., et al.\ 2018, \pasj, 70, S35.
        
        \bibitem[Medvedev et al.(2020)]{2020MNRAS.497.1842M} Medvedev, P., Sazonov, S., Gilfanov, M., et al.\ 2020, \mnras, 497, 1842.
        
        \bibitem[Momjian et al.(2008)]{2008AJ....136..344M} Momjian, E., Carilli, C.~L., \& McGreer, I.~D.\ 2008, \aj, 136, 344.
        
        \bibitem[Momjian et al.(2021)]{2021AJ....161..207M} Momjian, E., Ba{\~n}ados, E., Carilli, C.~L., Walter, F., \& Mazzucchelli, C.\ 2021, \aj, 161, 207.
        
        \bibitem[Mortlock et al.(2011)]{2011Natur.474..616M} Mortlock, D.~J., Warren, S.~J., Venemans, B.~P., et al.\ 2011, \nat, 474, 616.
        
        \bibitem[Narayan et al.(2003)]{2003PASJ...55L..69N} Narayan, R., Igumenshchev, I.~V., \& Abramowicz, M.~A.\ 2003, \pasj, 55, L69
        
        \bibitem[O'Dea \& Saikia(2021)]{2021A&ARv..29....3O} O'Dea, C.~P., \& Saikia, D.~J.\ 2021, \aapr, 29, 3.
        
        \bibitem[Pearson \& Readhead(1984)]{1984ARA&A..22...97P} Pearson, T.~J. \& Readhead, A.~C.~S.\ 1984, \araa, 22, 97.
        
        \bibitem[Perger et al.(2017)]{2017FrASS...4....9P} Perger, K., Frey, S., Gab{\'a}nyi, K. {\'E}., \& T{\'o}th, L.~V.\ 2017, Frontiers in Astronomy and Space Sciences, 4, 9.
        
        \bibitem[Perger et al.(2019)]{2019MNRAS.490.2542P} Perger, K., Frey, S., Gab{\'a}nyi, K. {\'E}., \& T{\'o}th, L.~V.\ 2019, \mnras, 490, 2542.
        
        \bibitem[Pradel et al.(2006)]{2006A&A...452.1099P} Pradel, N., Charlot, P., \& Lestrade, J.-F.\ 2006, \aap, 452, 1099.
        
        \bibitem[Ross \& Cross(2020)]{2020MNRAS.494..789R} Ross, N.~P., \& Cross, N.~J.~G.\ 2020, \mnras, 494, 789.
        
        \bibitem[Shepherd et al.(1994)]{1994BAAS...26..987S} Shepherd, M.~C., Pearson, T.~J., \& Taylor, G.~B.\ 1994, \baas, 26, 987.
        
        \bibitem[Spingola et al.(2020)]{2020A&A...643L..12S} Spingola, C., Dallacasa, D., Belladitta, S., et al.\ 2020, \aap, 643, L12.
        
        \bibitem[Tingay et al.(2013)]{2013PASA...30....7T} Tingay, S.~J., Goeke, R., Bowman, J.~D., et al.\ 2013, \pasa, 30, e007. 
        
        \bibitem[Wang et al.(2021)]{2021ApJ...908...53W} Wang, F., Fan, X., Yang, J., et al.\ 2021, \apj, 908, 53.
        
        \bibitem[Wang et al.(2021)]{2021ApJ...907L...1W} Wang, F., Yang, J., Fan, X., et al.\ 2021, \apjl, 907, L1.
        
        \bibitem[Wang et al.(2017)]{2017ApJ...835L..20W} Wang, R., Momjian, E., Carilli, C.~L., et al.\ 2017, \apjl, 835, L20.
        
        \bibitem[Wright(2006)]{2006PASP..118.1711W} Wright, E.~L.\ 2006, \pasp, 118, 1711.
        
        \bibitem[Wu et al.(2015)]{2015Natur.518..512W} Wu, X.-B., Wang, F., Fan, X., et al.\ 2015, \nat, 518, 512.
        
        \bibitem[Yang et al.(2020)]{2020ApJ...897L..14Y} Yang, J., Wang, F., Fan, X., et al.\ 2020, \apjl, 897, L14.
        
        
        
\end{thebibliography}
\def\nat{Nat.}
\def\apj{Astrophys. J.}
\def\aj{Astron. J.}
\def\apjl{Astrophys. J. Lett.}
\def\apjs{Astrophys. J. Suppl. Ser.}
\def\prd{Phys. Rev. D}
\def\pasp{Publ. Astron. Soc. Pacific}
\def\araa{Ann. Rev. Astron. Astrophys.}
\def\mnras{Mon. Not. R. Astron. Soc.}
\def\physrep{Phys. Rep.}
\def\aap{Astron. Astrophys.}
\def\araa{Annu. Rev. Astron. Astrophys.}
\def\pasa{Publ. Astron. Soc. Australia}
\def\cjaas{Chin. J. Astron. Astrophys Suppl.}
\def\baas{Bull. Amer. Astron. Soc.}
\def\pasj{Publ. Astron. Soc. Japan}
\def\ssr{Space Sci. Rev.}



%
%
\end{document}